\documentclass[aps, twocolumn, prl, longbibliography]{revtex4-1}
\usepackage{amsfonts}
\usepackage{amsmath}
\usepackage{amssymb}
\usepackage{mathptmx} % use Times font instead of Computer Modern
\usepackage{graphicx}
\usepackage{bm}
\usepackage{color}
\usepackage[colorlinks=true,linkcolor=blue,anchorcolor=red,citecolor=blue,urlcolor=blue]{hyperref}
\usepackage{ulem}
\usepackage{natbib}

\begin{document}
\title{Impact of in-situ controlled disorder screening on fractional quantum Hall effects and composite-fermion transport}
\author{T. Akiho and K. Muraki}
\affiliation{NTT Basic Research Laboratories, NTT Corporation, 3-1 Morinosato-Wakamiya, Atsugi 243-0198, Japan}
\keywords{one two three}
\pacs{PACS number}
\date{\today}

\begin{abstract}
We examine the impact of random potential due to remote impurites (RIs) and its in-situ controlled screening on fractional quantum Hall effects (FQHEs) around Landau-level filling factor $\nu = 1/2$. The experiment is made possible by using a dual-gate GaAs quantum well (QW) that allows for the independent control of the density $n_{e}$ of the two-dimensional electron system in the QW and that ($n_\text{SL}$) of excess electrons in the modulation-doping superlattice. As the screening is reduced by decreasing $n_\text{SL}$ at a fixed $n_{e}$, we observe a decrease in the apparent energy gap of the FQHEs deduced from thermal activation, which signifies a corresponding increase in the disorder broadening $\Gamma$ of composite fermions (CFs). Interestingly, the increase in $\Gamma$ is accompanied by a noticeable increase in the longitudinal resistivity at $\nu = 1/2$ ($\rho_{1/2}$), with a much stronger correlation with $\Gamma$ than electron mobility $\mu$ has. The in-situ control of RI screening enables us to disentangle the contributions of RIs and background impurities (BIs) to $\rho_{1/2}$, with the latter in good agreement with the CF theory. We construct a scaling plot that helps in estimating the BI contribution to $\rho_{1/2}$ for a given set of $n_{e}$ and $\mu$.
\end{abstract}
\maketitle

The fractional quantum Hall effect (FQHE)~\cite{Tsui1982} that clean two-dimensional electron systems (2DESs) exhibit in a strong perpendicular magnetic field ($B$) at low temperatures is a quintessential example of many-body topological phase and is thus attracting interest for the rich physics contained~\cite{Moore1991,Wen1995,Feldman2021} and also as a building block for fault-tolerant topological quantum computation~\cite{Nayak2008,Mong2014}.
FQHEs can be understood, both intuitively and quantitatively, by the composite-fermion (CF) theory~\cite{Jain1989}, which maps FQHEs to integer quantum Hall effects of a CF, an electron with an even number of flux quanta attached. The theory explains in a clear way the Landau-level filling factor $\nu$ ($= h n_{e}/eB$) at which FQHEs develop around $\nu = 1/2$ and their relative strength~\cite{Du1993,Manoharan1994} ($h$ is Planck's constant, $n_{e}$ is the electron density, and $e$ is the elementary charge). At the same time, the CF model maps the system of strongly interacting electrons at $\nu = 1/2$ to that of weakly interacting CFs at zero effective magnetic field, providing a theoretical framework to study the scattering of CFs and calculate the resistivity at $\nu = 1/2$~\cite{Halperin1993}.

While FQHEs have been observed in various material systems~\cite{DePoortere2002,Chung2018,Lai2004,Du2009,Bolotin2009,Dean2011,Polshyn2018,Zeng2019,Tsukazaki2010,Falson2015,Piot2010,Betthausen2014,Kott2014,Shi2015,Mironov2016,Ma2017,Komatsu2021,Shi2020}, GaAs remains the platform where the cleanest 2DESs can be materialized~\cite{Pfeiffer2003,Umansky2009,Chung2020}. In typical 2DESs formed in modulation-doped GaAs/AlGaAs heterostructures or quantum wells (QWs), two main scattering sources limit electron mobility $\mu$: background ionized impurities and remote ionized impurities, the latter introduced by modulation doping~\cite{Hwang2008}. As GaAs samples used for FQHE studies generally have thick spacer layers, which separate the 2DES from remote impurities (RIs), $\mu$ primarily reflects the background impurity (BI) concentration and has been used as the quality indicator of 2DESs. However, recent studies have shown that, in addition to improving $\mu$ by reducing BIs, properly screening the random potentials arising from RIs is mandatory for observing fragile FQHEs with a small energy gap~\cite{Umansky2009,Gamez2013}. Superlattice (SL) doping~\cite{Friedland1996,Chung2020PRM}, with Si donors incorporated in a narrow GaAs layer flanked by thin AlAs layers, is an effective way to implement this. Some of the doped electrons occupy the $X$ valleys of AlAs, where these ``excess'' electrons remain mobile and screen the RI potential without causing unwanted parallel conduction~\cite{Sammon2018, Sammon2018a}. Recently, we demonstrated the effect of RI screening on $\mu$ by controlling the excess electron density in-situ using a gate~\cite{Akiho2021}.
This suggests that the same technique can be used to study the impact of RI screening on FQHEs and CF transport.

In this paper, we study the impact of disorder and its screening on FQHEs in a GaAs 2DES by controlling in-situ the strength of RI screening. We measure the energy gap $\Delta_\nu$ of several FQHEs at $\nu = p/(2p \pm 1)$ ($p$ is an integer) around $\nu = 1/2$ and the resistivity at $\nu = 1/2$ ($\rho_{1/2}$) under different screening conditions. We observe that $\rho_{1/2}$ as well as $\Delta_\nu$ vary with the degree of screening. We extract the disorder broadening $\Gamma$ of CFs from the measured $\Delta_\nu$ and find that it is much more strongly correlated with $\rho_{1/2}$ than 2DES mobility $\mu$ is, indicating that the former is a better quality indicator for FQHEs. With the in-situ control of RI screening, we are able to disentangle the contributions of RIs and BIs to $\rho_{1/2}$. We use the CF theory to calculate the contribution of BIs to $\rho_{1/2}$ to find a good agreement with experiment. Based on these results, we construct a scaling plot, which allows one to estimate the contribution of BIs to $\rho_{1/2}$.

\begin{figure}[t]
\includegraphics[scale=1.08]{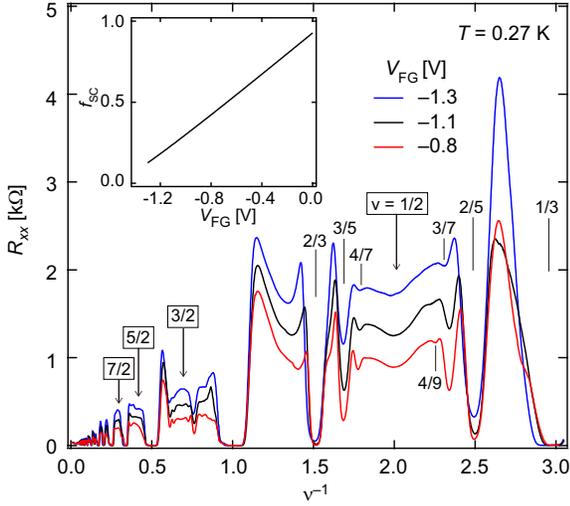}\caption{$R_{xx}$ vs. $\nu^{-1}$ at fixed density $n_{e} = 1.2 \times 10^{15}$ m$^{-2}$ with different $V_\text{FG}$ at $0.27$~K. $B$ was swept from 0 to 15 T. Red, black, and blue traces were taken under strong, intermediate, and weak screening conditions, respectively. Inset: $V_\text{FG}$ dependence of screening effect defined by $f_\text{sc} = n_\text{SL}/N_\text{Si}$.
}
\label{Fig1}
\end{figure}

The sample consists of a 30-nm-wide GaAs QW sandwiched between Al$_{0.27}$Ga$_{0.73}$As barriers, grown on an $n$-type GaAs (001) substrate. The QW, with its center located 207~nm below the surface, is modulation-doped on one side, with Si $\delta$-doping ($N_\text{Si} = 1 \times 10^{16}$~m$^{-2}$) at the center of the AlAs/GaAs/AlAs (2 nm/3 nm/2 nm) SL located 75~nm above the QW. The wafer was processed into a 100-$\mu$m-wide Hall bar with voltage probe distance of 120 $\mu$m and fitted with a Ti/Au front gate. The $n$-type substrate was used as a back gate. We measured FQHEs under different degrees of disorder screening by first setting the front-gate voltage ($V_\text{FG}$) at 4.3~K and waiting long enough for $n_{e}$ to stabilize before cooling the sample to 0.27~K. After the sample had cooled, we applied a back-gate voltage to adjust $n_{e}$ to the desired value. We use the quantity $f_\text{sc} = n_\text{SL}/N_\text{Si}$ as the parameter representing the degree of screening. Since $n_\text{SL}$ is not directly measurable, we estimated $n_\text{SL}$ and hence $f_\text{sc}$ by analyzing the $V_\text{FG}$ dependence of $n_{e}$ at 1.6~K.
The estimated $f_\text{sc}$ varies almost linearly with $V_\text{FG}$, as shown in the inset of Fig.~\ref{Fig1}.
More details of the estimation of $f_\text{sc}$ are described in Ref.~\cite{Akiho2021}.

Figure~\ref{Fig1} shows the longitudinal resistance $R_{xx}$ measured at a fixed density of $n_{e} = 1.2\times 10^{15}$~m$^{-2}$ with $V_\text{FG} = -0.8$, $-1.1$, and $-1.3$ V, corresponding to strong, intermediate, and weak screening ($f_\text{sc} = 0.42$, $0.24$, and $0.13$), respectively, plotted as a function of $\nu^{-1}$. The FQHEs at $\nu = 1/3$ and $2/3$ are nearly fully developed under all conditions. On the other hand, those at $\nu = 2/5$, $3/5$, $3/7$, and $4/7$ clearly become weaker as $V_\text{FG}$ is lowered, and hence the screening is reduced. In addition, the dip at $\nu = 4/9$, which is visible under the strong screening condition, disappears under the weak screening condition. Around $\nu = 3/2$, similar $V_\text{FG}$ dependence is seen for FQHEs at $\nu = 4/3$ and $5/3$.

\begin{figure}[t]
\includegraphics[scale=1.08]{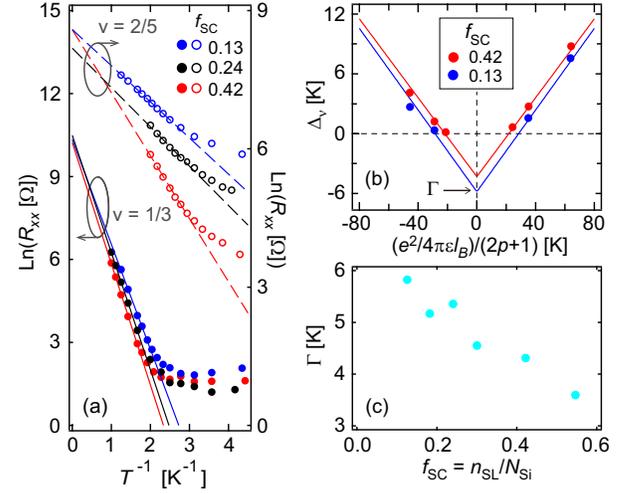}\caption{(a) Arrhenius plot for $R_{xx}$ at $\nu = 1/3$ (filled circles) and $2/5$ (open circles) under different screening conditions. (b) Scaling of activation energy for FQHEs around $\nu = 1/2$ under different screening conditions, deduced from the Arrhenius plot. The solid lines represent fits using Eq.~(1). The magnitude of the negative intercept of these lines with the $y$ axis gives $\Gamma$. (c) $\Gamma$ obtained from the scaling law fitting, plotted as a function of $f_\text{sc}$.
}
\label{Fig2}
\end{figure}

To characterize the impact of the screening on FQHEs quantitatively, we measured the temperature ($T$) dependence of $R_{xx}$ and deduced the energy gap $\Delta_{\nu}$.
Figure~\ref{Fig2}(a) plots ln($R_{xx}$) vs $1/T$ at $\nu = 1/3$ and $2/5$, corresponding to $p = 1$ and $2$, measured under three different screening conditions. We obtain $\Delta_{\nu}$ by fitting the data in the temperature range where the activated behavior is seen with $R_{xx} \propto \exp(-\Delta_{\nu}/2T)$.
In the same way, we also estimated $\Delta_{\nu}$ for FQHEs at $\nu = 2/3$, $3/5$, $4/7$, and $3/7$  (corresponding to $p = -2$, $-3$, $-4$, and $3$, respectively) under different screening conditions. To systematically analyze the obtained $\Delta_{\nu}$ for different $p$'s, we used the scaling law introduced in Ref.~\cite{Manoharan1994}:
\begin{equation}
	\Delta_{\nu} = \frac{\kappa}{|2p+1|} \frac{e^{2}}{4\pi\epsilon \ell_{B}} - \Gamma
\end{equation}
where $\epsilon = \epsilon_{r}\epsilon_{0}$ with $\epsilon_{0}$ the vacuum permittivity and $\epsilon_{r} = 13$ for (Al)GaAs, $\ell_{B} = (\hbar/eB)^{1/2}$ is the magnetic length with $\hbar = h/2\pi$, $\kappa$ is a dimensionless parameter representing the strength of the Coulomb interaction, and $\Gamma$ denotes the gap reduction due to disorder. By plotting $\Delta_{\nu}$'s for different $p$'s as a function of $(e^{2}/4\pi\epsilon\ell_{B})/(2p + 1)$ [Fig.~\ref{Fig2}(b)] and fitting them using Eq.~(1), $\kappa$ and $\Gamma$ are obtained from the slope and intercept, respectively. The data for the weak and strong screening can be fitted using the same $\kappa$ value ($= 0.197 \pm 0.007$), indicating that the excess electrons in the SL do not discernibly affect the strength of the intralayer Coulomb interaction responsible for the FQHEs. In contrast, the impact on $\Gamma$ is obvious---$\Gamma$ decreasing upon increasing screening. Measurements for various $f_\text{sc}$ values, summarized in Fig.~\ref{Fig2}(c), reveal that $\Gamma$ increases from 3.6 to 5.8 K as $f_\text{sc}$ decreases from 0.55 to 0.13. As $\Gamma$ can be viewed as representing the Landau-level broadening for CFs, these results confirm that the in-situ control of the visibility of the FQHEs demonstrated in Fig.~\ref{Fig1}(a) is due to the controlled screening of disorder.

\begin{figure}[ptb]
\includegraphics[scale=1.08]{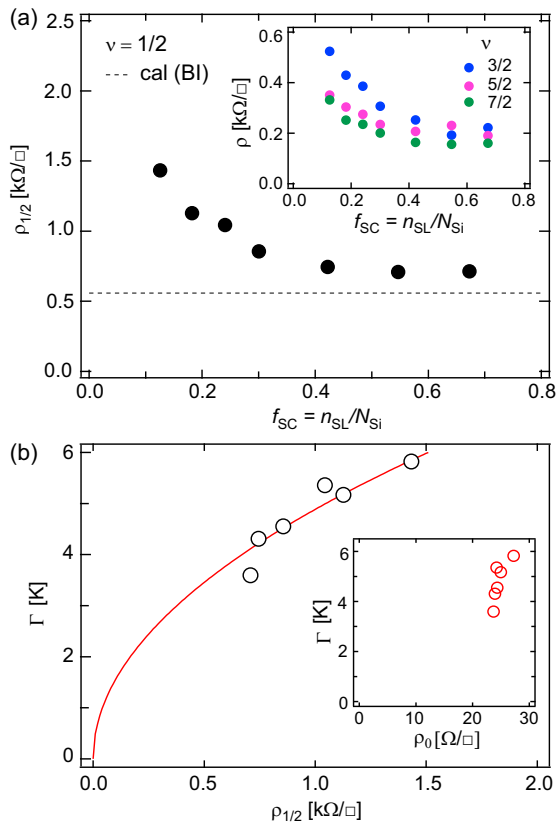}\caption{(a) $\rho_{1/2}$ under different screening conditions at $n_{e} = 1.2 \times 10^{15}$~m$^{-2}$. The dashed line shows the contribution of BI scattering to $\rho_{1/2}$ calculated using Eq.~(2) modified for BIs. Inset: Resistivity at $\nu = 3/2$, $5/2$, and $7/2$ under different screening conditions. (b) Plots of $\Gamma$ with respect to $\rho_{1/2}$ under different screening conditions. The solid line indicates as a guide the fitting using the $(\rho_{1/2})^{0.5}$ function. Inset: Plots of $\Gamma$ with respect to the zero field resistivity $\rho_{0}$.
}
\label{Fig3}
\end{figure}

Another important observation in Fig.~\ref{Fig1} is that, with decreasing $f_\text{sc}$, $R_{xx}$ increases not only in the FQHE regions but also in regions between them. We focus on the state at $\nu = 1/2$ and plot $\rho_{1/2}$ as a function of $f_\text{sc}$ in Fig.~\ref{Fig3}(a). $\rho_{1/2}$ increases noticeably with decreasing $f_\text{sc}$ below $0.42$ ($V_\text{FG} < -0.8$~V), whereas it is almost constant for $f_\text{sc} \geq 0.42$. Similar $f_\text{sc}$ dependences are observed for other half-integer fillings $\nu = 3/2$, $5/2$, and $7/2$ [inset of Fig.~\ref{Fig3}(a)]. To examine the correlation between FQHEs and CF transport, we plot $\Gamma$ versus $\rho_{1/2}$ in Fig.~\ref{Fig3}(b). Their relation can be fitted approximately by $\Gamma \propto \rho_{1/2}^{0.5} $ as shown by the solid line. For comparison, we plot $\Gamma$ against $\rho_{0}$, the resistivity at zero magnetic field, a quantity directly related to $\mu$ ($= 1/e n_{e} \rho_{0}$) [inset of Fig.~\ref{Fig3}(b)]. When $\Gamma$ varies by 38\%, $\rho_{0}$ only changes by 13\% ($\mu=191$--$219$~m$^{2}$/Vs), whereas $\rho_{1/2}$ changes by 51\% for the same $\Gamma$ range, demonstrating that $\rho_{1/2}$ is more strongly correlated with the visibility of FQHEs. This result  bears an intriguing similarity with the recent report that the resistivity at $\nu = 5/2$ in the high-temperatures regime serves as an indicator of the strength of the $\nu = 5/2$ FQHE that emerges at low temperatures~\cite{Qian2017a}.

Now we discuss the scattering mechanism that determines $\rho_{1/2}$. In the CF model, at $\nu = 1/2$ CFs experience a zero effective magnetic field in the mean field and form a Fermi surface. According to Ref.~\cite{Halperin1993}, scattering of CFs at $\nu = 1/2$ is dominated by fluctuations in the electron density induced by the charged impurities randomly distributed in the modulation-doped layer, which translate into fluctuations in the Chern-Simons gauge field and act as random magnetic fields with zero mean. The $\rho_{1/2}$ due to this scattering mechanism is given by \cite{Halperin1993}
\begin{equation}
	\rho_{1/2} = \frac{n_\text{imp}}{n_{e}} \frac{1}{k_{F} d_{s}} \frac{4\pi \hbar}{e^{2}},
\end{equation}
where $n_\text{imp}$ is the sheet density of the ionized impurities, $k_{F} = (4 \pi n_{e})^{1/2}$ is the Fermi wave number of spin-polarized CFs, and $d_{s}$ is the distance between the 2DES and the doped layer. In the ideal case of no BIs or charge traps, we have $n_\text{imp} = n_{e}$, where $\rho_{1/2}$ takes a minimum value determined solely by the factor $k_{F}d_{s}$ ($= d_{s}/\ell_{B}$ at $\nu = 1/2$).

In our sample, the density of excess electrons in the SL doping layer can be varied via $V_\text{FG}$. This can be thought of as effectively varying $n_\text{imp}$ in Eq.~(2), which enables us to disentangle the contribution of RIs to $\rho_{1/2}$ from other ones. In Fig.~\ref{Fig3}(a), $\rho_{1/2}$ first decreases with increasing $f_\text{sc}$ and then becomes almost constant for $f_\text{sc} > 0.4$. Thus, we can clearly identify the increase in $\rho_{1/2}$ at $f_\text{sc} < 0.4$ as due to RI scattering. On the other hand, this suggests that at $f_\text{sc} > 0.4$ the screening is sufficient to make the contribution of RIs insignificant. To examine the mechanism that determines $\rho_{1/2}$ in this well-screened regime, we estimated the contribution of BIs by modifying Eq.~(2). We replaced $n_\text{imp}$ and $d_{s}$ in Eq.~(2) with $n_\text{BI}(z)dz$, the sheet density of BIs within a slice $dz$ at each position $z$ along the growth direction, and $\langle d(z)\rangle$, the expectation value of the distance from that position to the 2DES, respectively, and integrated Eq.~(2) over $z$
\footnote{The divergence at $d_{s} = 0$ that occurs when applying Eq.~(2) to BIs can be avoided by taking into account the finite thickness of the 2DES. Using the wave function $\psi(z)$ of the 2DES, we replaced $d_{s}$ in Eq.~(2) with $\langle d(z)\rangle = \int |z - z^\prime||\psi(z^\prime)|^2 dz^\prime$, which remains finite for all $z$. When the BI concentration ($n_\text{BI}$) is constant, the only $z$-dependent factor in the modified Eq.~(2) is $\langle d(z)\rangle^{-1}$, with its integration $\beta \equiv \int \langle d(z)\rangle^{-1}dz$ being a dimensionless number. Thus, $\rho_{1/2}$ due to BI scattering is given by $\rho_{1/2} = (\beta/\pi^{1/2})(n_\text{BI}/n_{e}^{3/2})(h/e^{2})$. For a 30-nm-thick QW with infinite barrier, we have $\beta = 9.4$.
}.
A calculation using a constant $n_\text{BI}$ of $1.7 \times 10^{14}$~cm$^{-3}$, deduced from the analysis of mobility, gives $\rho_{1/2} = 0.56$~k$\Omega/\square$ [shown by the horizontal dashed line in Fig.~3(a)], which accounts for the $\rho_{1/2}$ values at $f_\text{sc} > 0.4$ surprisingly well.

\begin{figure}[tpb]
\includegraphics[scale=1.08]{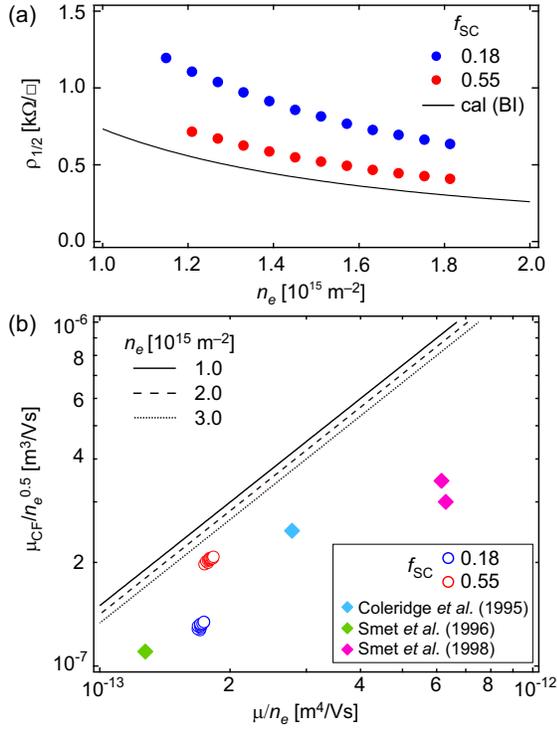}\caption{(a) $n_{e}$ dependence of $\rho_{1/2}$. Red and blue symbols are data taken under strong and weak screening conditions, respectively. The solid curve is a calculation of BI contribution to $\rho_{1/2}$. (b) Scaling plot of CF and 2DES mobilities normalized by $n_{e}^{0.5}$ and $n_{e}$, respectively. Circles are replots of the data in (a) taken under strong (red) and weak (blue) screening conditions. As a reference, other reported values are also plotted (diamonds). The solid, dashed, and dotted lines are the calculation results for $n_{e} = 1.0$, $2.0$, and $3.0 \times 10^{15}$~m$^{-2}$, respectively, obtained with varying $n_\text{BI}$.
}
\label{Fig4}
\end{figure}

Next, we quantitatively investigate the contribution of RIs to $\rho_{1/2}$ and the impact of controlled screening therein. We examined the $n_{e}$ dependence of $\rho_{1/2}$ by varying $n_{e}$ with the back gate at a fixed $V_\text{FG}$. Note that the back gate barely affects $n_\text{SL}$, which ensures that the screening condition remains nearly constant upon varying $n_{e}$. The results for $V_\text{FG} = -1.2$ and $-0.6$ V, which correspond to the weak ($f_\text{sc} = 0.18$) and strong ($f_\text{sc} = 0.55$) screening, respectively, are shown in Fig.~\ref{Fig4}(a). The solid line indicates the calculated $\rho_{1/2}$ due to BI scattering, assuming the same $n_\text{BI}$ as above. The calculation well accounts for the data for $f_\text{sc} = 0.55$, consistent with the expected $n_{e}^{-3/2}$ dependence, which corroborates that in our sample $\rho_{1/2}$ in the well-screened regime is dominated by BI scattering. On the other hand, we are able to unambiguously ascribe the difference between the $\rho_{1/2}$ values for $f_\text{sc} = 0.18$ and $0.55$ to RI scattering. We find that the difference can be well fitted by Eq.~(2). Taking $d_{s}$ to be the center-to-center distance $90$~nm between the QW and the doping SL, we obtain $n_\text{imp} = 1.0 \times 10^{14}$~m$^{-2}$ from the fit. We note that this is only 2.7\% that of the difference in the remote ionized impurity density if we simply evaluate it as $N_\text{Si} - n_\text{SL} = (1-f_\text{sc})N_\text{Si}$. Although it is known that Eq.~(2) tends to overestimate $\rho_{1/2}$ compared to experimental values for high-quality 2DESs~\footnote{It is not clear why Eq.~(2) tends to overestimate $\rho_{1/2}$ due to RIs when it yields reasonable results when modified for BIs. Possible reasons include spatial correlation among remote ionized impurities and screening due to excess electrons generated by photo illumination~\cite{Gamez2013}} (e.g., by a factor of $\sim 3$~\cite{Halperin1993,Du1993}), the above reduction factor of 2.7\%  is much more significant. It indicates that the screening by the excess electrons is effective even in the weak screening case, making the simple analysis regarding RIs as an ensemble of unscreened charges inadequate, similarly to what has been reported for 2DES mobility~\cite{Akiho2021, Sammon2018a,Sammon2018}.

Finally,  we examine the relation between CF mobility $\mu_\text{CF} = 1/en_{e}\rho_{1/2}$ and 2DES mobility $\mu$. As we have shown, $\mu_\text{CF}$ is governed by both RIs and BIs, whereas in typical high-mobility GaAs 2DESs with large $d_{s}$, $\mu$ is governed mostly by BIs~\cite{Hwang2008}. This suggests that one can take $\mu$ as a measure of $n_\text{BI}$ and use this $n_\text{BI}$ to estimate $\mu_\text{CF}$ limited by BIs. Then, deviation of measured $\mu_\text{CF}$ from this value can be ascribed to RI scattering. To do this at one go for different densities, we construct a scaling plot as follows. As $k_{F} \propto n_{e}^{1/2}$ in Eq.~(2), we have $\rho_{1/2} \propto n_{e}^{-3/2}$ and hence $\mu_\text{CF} \propto n_{e}^{1/2}$. For GaAs 2DESs, it is known that the approximate relation $\mu \propto n_{e}^{\alpha}$ holds for BI-limited mobility, with $\alpha \approx 1$ for $n_{e}=1$--$2 \times 10^{15}$~m$^{-2}$~\cite{Hwang2008}. We therefore make a plot of $\mu_\text{CF}/n_{e}^{1/2}$ versus $\mu/n_{e}$ as shown in Fig.~\ref{Fig4}(b), where we plot the experimental data in Fig.~\ref{Fig4}(a) together with calculations for several densities obtained with varying $n_\text{BI}$. Data in the literature for GaAs 2DESs with conventional modulation doping, with both $\rho_{1/2}$ and $\mu$ available~\cite{Coleridge1995, Smet1996, Smet1998}, are also plotted for comparison. Due to the scaling, the calculated curves for different densities are placed close to each other. Similarly, our data for various $n_{s}$ concentrate around two points for the weak and strong screening. We observe that all the experimental data lie below the calculated curves, which suggests the influence of RI scattering. Among all the data plotted here, our data for the strong screening lie closest to the calculated curves, indicating efficient screening of RIs. This is reasonable, as the samples in the literature employed conventional modulation doping. It would therefore be interesting to add data for recent ultrahigh-quality samples with SL doping~\cite{Chung2020} to this plot, which will be possible if $\rho_{1/2}$ is available. We believe that our analysis and the basic idea of the scaling plot are helpful in identifying the mechanisms limiting the visibility of FQHEs and improving sample design and growth of various materials not limited to modulation-doped GaAs QWs or heterostructures.

In summary, we investigated the impact of in-situ controlled disorder screening on FQHEs. We found that the screening of RIs impacts not only the visibility of the FQHEs but also $\rho_{1/2}$, the resistivity at $\nu = 1/2$, or CF mobility. In the well-screened regime, the measured $\rho_{1/2}$ agrees well with that due to BIs estimated using the CF theory. The strong correlation between the strength of FQHEs and $\rho_{1/2}$ proves $\rho_{1/2}$, or CF mobility, to be a better quality indicator for FQHEs than 2DES mobility.

The authors thank Michiyo Kamiya for measurement support and Hiroaki Murofushi for processing the device. This work was supported by JSPS KAKENHI Grant No. JP15H05854.

%\bibliography{ref_takm0314}

%\begin{thebibliography}{37}%
\end{document}